\documentclass[prl,twocolumn,aps,letterpaper, preprintnumbers,superscriptaddress]{revtex4}
\usepackage{upgreek}
\usepackage{hyperref}
\usepackage{verbatim}
\usepackage{amsmath}
\usepackage{latexsym}
\usepackage{revsymb}
\usepackage{yfonts}
\usepackage{ifthen}
\usepackage{tcolorbox}
\usepackage{natbib}
\usepackage{amsfonts}
\usepackage{amsmath}
\usepackage{amssymb}
\usepackage{amsthm}
\usepackage{graphicx}
\usepackage{bm}
\usepackage{bbm}
\usepackage{epsfig,color,amssymb}
\usepackage{subfigure}
\usepackage{amsfonts}
\usepackage{amscd}
\usepackage{amsmath}
\usepackage{multirow}
\usepackage{chemarrow}
\usepackage{dcolumn}
\usepackage{bm}
\usepackage{graphicx}
\usepackage{enumerate}
\usepackage{epsfig}
\usepackage{subfigure}
\usepackage{xcolor}
\usepackage{multirow}
\usepackage{ulem}
\usepackage{braket}
\usepackage{comment}
\usepackage{enumitem}
\usepackage{amsthm}
\usepackage{diagbox}

\renewcommand{\emph}[1]{\textit{#1}}

\usepackage{siunitx}

\newcommand{\ba}{\begin{array}}
\newcommand{\ea}{\end{array}}
\newcommand{\be}{\begin{equation}}
\newcommand{\ee}{\end{equation}}
\newcommand{\bea}{\begin{eqnarray}}
\newcommand{\eea}{\end{eqnarray}}

\begin{document}
\title{Optical skyrmion lattices accelerating in free space}
\author{Haijun Wu}
\affiliation{Wang Da-Heng Center, Heilongjiang Key Laboratory of Quantum Control, Harbin University of Science and Technology, Harbin 150080, China}
\affiliation{Centre for Disruptive Photonic Technologies, School of Physical and Mathematical Sciences, Nanyang Technological University, Singapore 637371, Singapore}

\author{Weijie Zhou}
\affiliation{Centre for Disruptive Photonic Technologies, School of Physical and Mathematical Sciences, Nanyang Technological University, Singapore 637371, Singapore}

\author{Zhihan Zhu}
\affiliation{Wang Da-Heng Center, Heilongjiang Key Laboratory of Quantum Control, Harbin University of Science and Technology, Harbin 150080, China}

\author{Yijie Shen}\email{yijie.shen@ntu.edu.sg}
\affiliation{Centre for Disruptive Photonic Technologies, School of Physical and Mathematical Sciences, Nanyang Technological University, Singapore 637371, Singapore}
\affiliation{School of Electrical and Electronic Engineering, Nanyang Technological University, Singapore 639798, Singapore}

\date{\today}


\begin{abstract}
Generation and propagation of optical skyrmions provide a versatile plalform for topologically nontrivial optical informatics and light-matter interactions, but their acceleration along curved trajectories is to be studied.
In this study, we experimentally demonstrate the first accelerating skyrmion lattices conveyed by Airy structured light, characterized by topologically stable skyrmion textures with self-acceleration along parabolic trajectories.
We show that the skyrmion unit cell can maintain a Skyrme number $|N_\text{sk}|>0.9$ within a propagation range of $\pm1.22\ z_R$ upon parabolic acceleration. Notably, the meron structure remains $|N_\text{sk}|$ stable within $0.5\pm0.02$ over a significantly extended range of $\pm3.06\ z_R$. Our work provides a new potential carrier for topologically robust information distribution, particle sorting and manipulation. 
\end{abstract}

\maketitle

\section{\label{sec:level1}Introduction}
Optical skyrmions are a kind of vectorial structured light possessing nontrivial particle-like topologies~\cite{shen2024optical}. Topological structures inherently bring protected stability in condensed matter physics~\cite{bernevig2022progress,bogdanov2020physical,fert2017magnetic,han2022high,chen2024all}, similar studies in optics are also emerging -- topologies can make structured light robust against external perturbations~\cite{wang2024topological,liu2022disorder,wang2024observation,ornelas2024non}. Unlike the condensed-matter-based skyrmions localized in materials for potential data storage or local data processing~\cite{bogdanov2020physical,fert2017magnetic,han2022high,chen2024all}, the optical skyrmions possess the ability to propagate in free space, which are invaluable for practical information transfer applications and nontrivial light-matter interaction~\cite{wan2023ultra,wang2024single,habibovic2024emerging,tamura2024direct}.
Recently, exploiting advanced technologies of structured light modulation~\cite{forbes2021structured,he2022towards,rosales2024perspective}, both individual skyrmions and skyrmion lattices can be not only generated~\cite{tsesses2018optical,du2019deep,davis2020ultrafast,dai2020plasmonic}, but also propagating in free space with controlled topologies~\cite{gao2020paraxial,shen2021generation,hakobyan2024unitary,he2024optical,kerridge2024optical,lin2024chip}.
Especially, it becomes a mainstream to extend more complex topological structures controlled in higher dimensions, such as conformal skyrmion and meron lattices~\cite{marco2024propagation,marco2024periodic}, bimerons~\cite{shen2021topological,bervskys2023accelerating}, multiskyrmions and multimerons~\cite{shen2024topologically,mcwilliam2023topological}, 
and hopfions~\cite{shen2023topological,sugic2021particle,tamura2024three}, with more flexible control and resilient propagation. However, all the experimentally observed free-space skyrmions can only propagate straightly. Controlling topological skyrmions that spatially accelerate along curved trajectories is crucial for higher-level information sorting or transfer but remains a challenge.


In this work, we address this gap by demonstrating the generation of accelerating skyrmion lattices in paraxial beams and investigating their topological properties during propagation. Specifically, we generate paraxial skyrmion lattices through the non-separable superposition of orthogonally polarized Airy and vortex Airy beams. Through theoretical simulations and experimental verification, we show that the periodic Gaussian- and vortex-like intensity profiles of these beams naturally form skyrmion structures, which then evolve into stable skyrmion lattices. Furthermore, we demonstrate that the skyrmion and meron lattices maintain robust topological properties during free-space propagation, with stability observed over propagation ranges of $\pm1.22\ z_R$ and $\pm3.06\ z_R$, respectively, confirming their resilience under the influence of propagation effects.
These findings highlight the potential of accelerating optical skyrmion lattices for future applications in topological information distribution, particle sorting and manipulation in nontrivial light-matter interactions, opening new directions in structured light research.

\section{Concept \& Principle}
\begin{figure*}[htbp]
\includegraphics[width=0.9\linewidth]{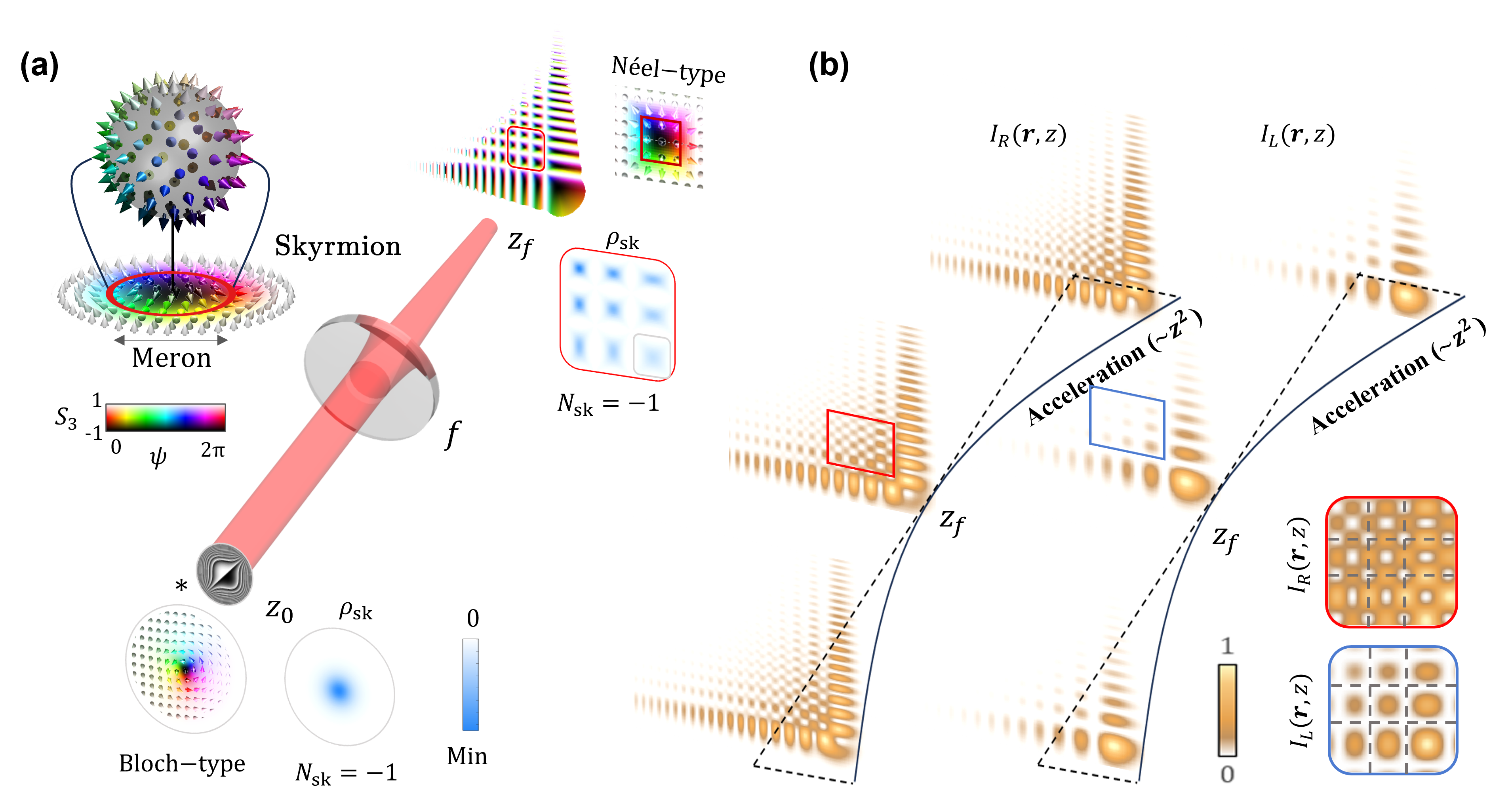}
\caption{\label{fig1} \textbf{Concept of periodically accelerating skyrmion lattices.} \textbf{a}, Schematic of periodic skyrmion lattices generation via cubic phase modulation. The original Bloch-type skyrmion texture ($N_{\mathrm{sk}}=-1$) transforms into square skyrmion lattices, each unit cell retaining $N_{\mathrm{sk}}=-1$. The color map represents the Stokes vector orientation mapped to HSL color space, with maximum saturation. Hue encodes transverse orientation $\psi= \arg(\mathrm{S}_1 + i \mathrm{S}_2)$, and lightness corresponds to $\mathrm{S}_3$. \textbf{b}, Airy and vortex Airy beams, exhibiting periodic intensity lattices which complementary to each other (highlighted in red and blue regions), accelerate along curved trajectories.}
\end{figure*}
To better understand the optical skyrmion lattices described in this study, we first revisit the fundamental theory of skyrmions. The classic N\'eel-type skyrmion texture, as depicted in the inset of Fig. 1(a), is formed by unwrapping the vectors of a 3D parametric sphere onto a 2D transverse plane, resulting in a skyrmion structure represented by a 3D vector distribution $\mathbf{S}(x,y)$. Similarly, unwrapping hemispheres of the sphere forms meron textures, also analyzed in this study. Generally, for both skyrmion or merion textures, the topological invariant (times of the vector $\mathbf{S}$ wraps around the sphere) is quantified by the Skyrme number~\cite{shen2024optical}:
\begin{equation}\label{Eq. (1)}
    N_{\mathrm{sk}} = \frac{1}{4\pi} \iint \mathbf{S} \cdot \left( \frac{\partial \mathbf{S}}{\partial x} \times \frac{\partial \mathbf{S}}{\partial y} \right) dx \, dy,
\end{equation}

\noindent where the integrand $\rho_{\mathrm{sk}}=\mathbf{S}\cdot\left(\partial_x\mathbf{S}\times\partial_y\mathbf{S}\right)$ represents the skyrme density.

Typically, optical skyrmions are created by the coherent combination of orthogonally polarized Laguerre-Gaussian (LG) beams with different topological charges~\cite{allen1992orbital}. The field distribution is given by:
\begin{equation}\label{Eq. (2)}
    \mathbf{E}(x, y) = \sqrt{\alpha} \, \mathrm{LG}_{\mathrm{\ell}_{1},0}(x, y) \, \hat{\mathbf{e}}_R + e^{i\theta} \sqrt{1 - \alpha} \, \mathrm{LG}_{\mathrm{\ell}_{2},0}(x, y) \, \hat{\mathbf{e}}_L
\end{equation}

\noindent where, $\alpha \in [0,1]$ defines the relative amplitude of the two modes, and $\mathrm{LG}_{\ell,p}$ represents the LG mode with azimuthal and radial indices $\ell$ and $p$, respectively (see \textbf{Supplementary Material S1} for detailed expressions). Here, $\left|\ell_1\right|\neq\left|\ell_2\right|$, and $e^{i\theta} $ determines the intramodal phase, which governs the skyrmion state, and ${\hat{\mathbf{e}}}_R$ and ${\hat{\mathbf{e}}}_L$ are unit vectors of right-circular polarization (RCP) and left-circular polarization (LCP). By modulating the parameters in Eq.~(\ref{Eq. (1)}), we can construct the distribution of normalized Stokes vector $\mathbf{S}(x,y)=\left[\mathrm{S}_1\left(x,y\right),\ \mathrm{S}_2\left(x,y\right),\mathrm{S}_3(x,y)\right]^T$, which describes the 3D vector distribution of optical topological textures~\cite{shen2021generation}. 

Returning to the main focus, A natural route to construct accelerating skyrmion lattices leverages the unique properties of Airy beams, which exhibit non-diffracting, self-accelerating intensity distributions with periodic lattice structures in the transverse plane~\cite{efremidis2019airy,shi2017interaction}. While ideal Airy beams require infinite energy, finite-energy Airy beams can be realized through the Fourier transform of a Gaussian beam modulated by a cubic phase. Introducing vortex phases during this process results in vortex Airy beams, which generate periodic vortex lattices complementary to the original Airy intensity distribution due to spiral phase contrast (see \textbf{Supplementary Material S2} for more details)~\cite{dai2010propagation,qiu2018spiral,pan2022airy}. Thus, the skyrmion lattices in this work can be described as:
\begin{equation}\label{Eq. (3)}
\begin{aligned}
\mathbf{E}_{\mathrm{T}}(x_0, y_0) = \mathcal{F} \left[ \mathbf{E}(x, y) \exp\left(\frac{i b^3 \left(x^3 + y^3\right)}{3}\right) \right] \\
= \sqrt{\alpha} A_{\ell_1}(x_0, y_0) \, \hat{\mathbf{e}}_R 
+ e^{i \theta} \sqrt{1 - \alpha} A_{\ell_2}(x_0, y_0) \, \hat{\mathbf{e}}_L
\end{aligned}
\end{equation}

\noindent Where, $\mathrm{A}_\ell\left(\cdot\right)$ is vortex Airy beam converted by topological charge $\ell$, and $b$ is the scaling factor of the Airy function determining the rate of transverse acceleration.
This process is shown as Fig.~\ref{fig1}(a), for $\ell_1=1$, $\ell_2=0$ and $\theta=-\pi/2$, the skyrmion texture gradually evolves into periodic skyrmion lattices at the Fourier plane, displaying square symmetry in their unit cells. Despite the geometric transformation, the Skyrme number for each cell remains $N_{\mathrm{sk}}=-1$, equivalent to the original circular skyrmions.

The transition between Bloch-type and N\'eel-type skyrmion textures at different propagation planes ($z_0$ to $z_f$) is attributed to the accumulation of Gouy phase, which varies with the spatial mode order~\cite{zhong2021gouy}. As shown in Fig.~\ref{fig1}(b), during propagation, both Airy and vortex Airy beams exhibit 3D intensity profiles that accelerate along parabolic trajectories, transferring their self-accelerating behavior to the constructed skyrmion lattices. This mechanism enables the skyrmion lattices to maintain their topological structures within a specific propagation range, ensuring stability and robustness.

\section{Experimental Setup}
\begin{figure}
\includegraphics[width=\linewidth]{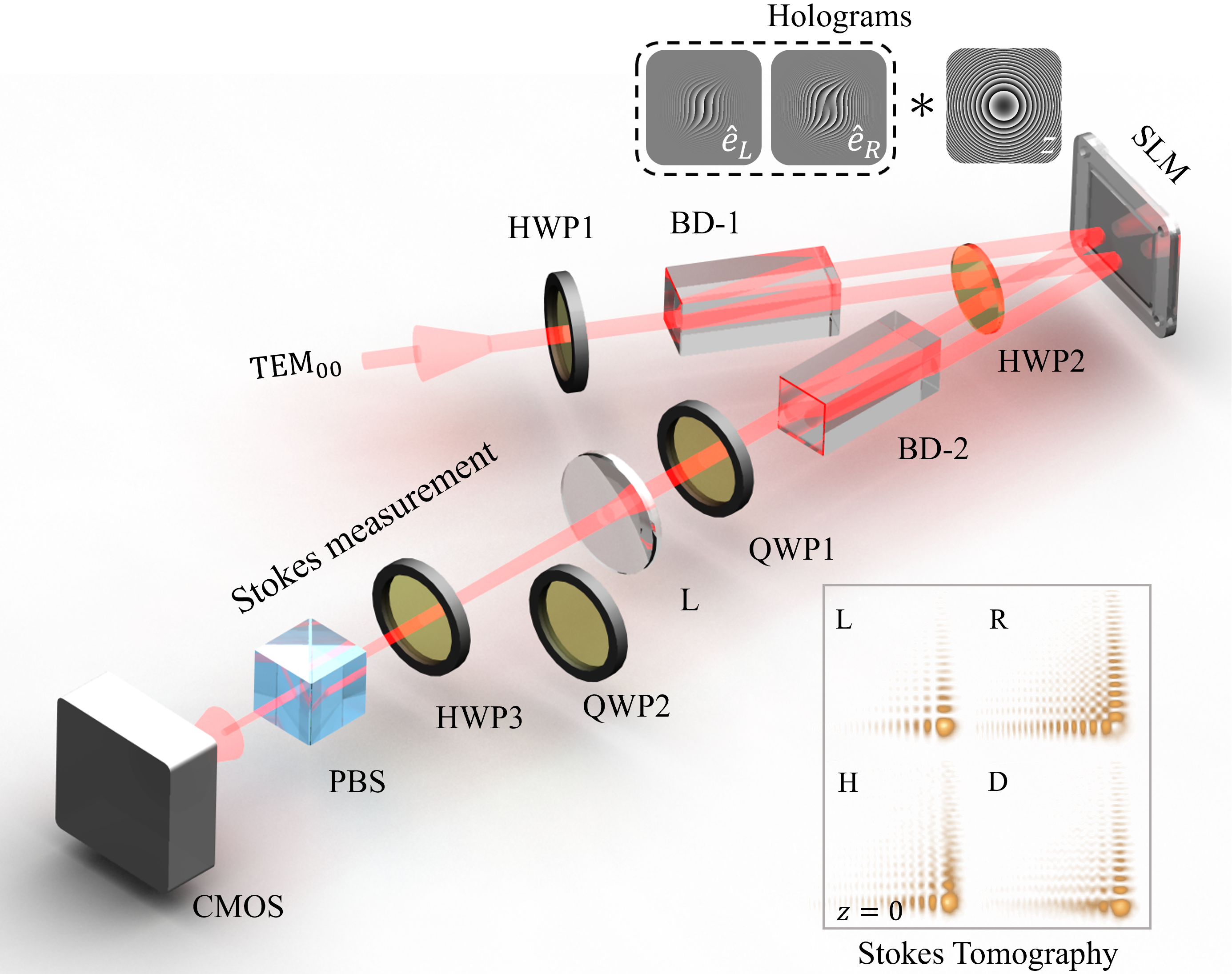}
\caption{\label{fig2} \textbf{Experimental setup.} The key components include, beam displacer prisms (BD), spatial light modulator (SLM), a camera (CMOS), half-wave plate (HWP), quarter-wave plate (QWP).}
\end{figure}

To further demonstrate the accelerating properties of skyrmion lattices, the experimental setup is depicted in Fig.~\ref{fig2}. A self-locking polarization Mach-Zehnder interferometer, incorporating two beam displacers (BD-1 and BD-2), forms the core of the system. A continuous 532 nm ${\rm TEM}_{00}$ laser beam (CNI Laser MSL-III-532) passes through BD-1, splitting into two orthogonally polarized beams directed to separate regions of a spatial light modulator (SLM, Holoeye ERIS-NIR-153). These beams are individually modulated using complex amplitude modulation (CAM) holograms corresponding to the Fourier transforms of the desired Airy and vortex Airy beams~\cite{rosales2017shape}. The beam radius $w_0$ and scaling factor \textit{b} are set as to 1~mm and $\sqrt[3]{3/11}$, respectively. A half-wave plate (HWP2, fixed at $45^\circ$) ensures horizontal polarization for both beams before modulation by the SLM.  After modulation, BD-2 recombines the beams. A quarter-wave plate (QWP1, fixed at $45^\circ$) and a Fourier lens (L) with a focal length of 200 mm are used to reconstruct the Airy vector beam at the Fourier plane. The propagation tomographies of the Airy beams are recorded via a digital propagation technique integrated into the CAM holograms, eliminating the need for physical translation of the CMOS camera (Allied Vision Alvium 1800 U-240m) along the $z$-axis~\cite{osten2014recent,schulze2012beam}. Furthermore, the camera, combined with polarizers (QWP2, HWP3, and PBS), performs spatial Stokes tomography, capturing the structures of skyrmion lattices in Stokes space.

\section{Results \& Discussion}
\begin{figure}
\includegraphics[width=\linewidth]{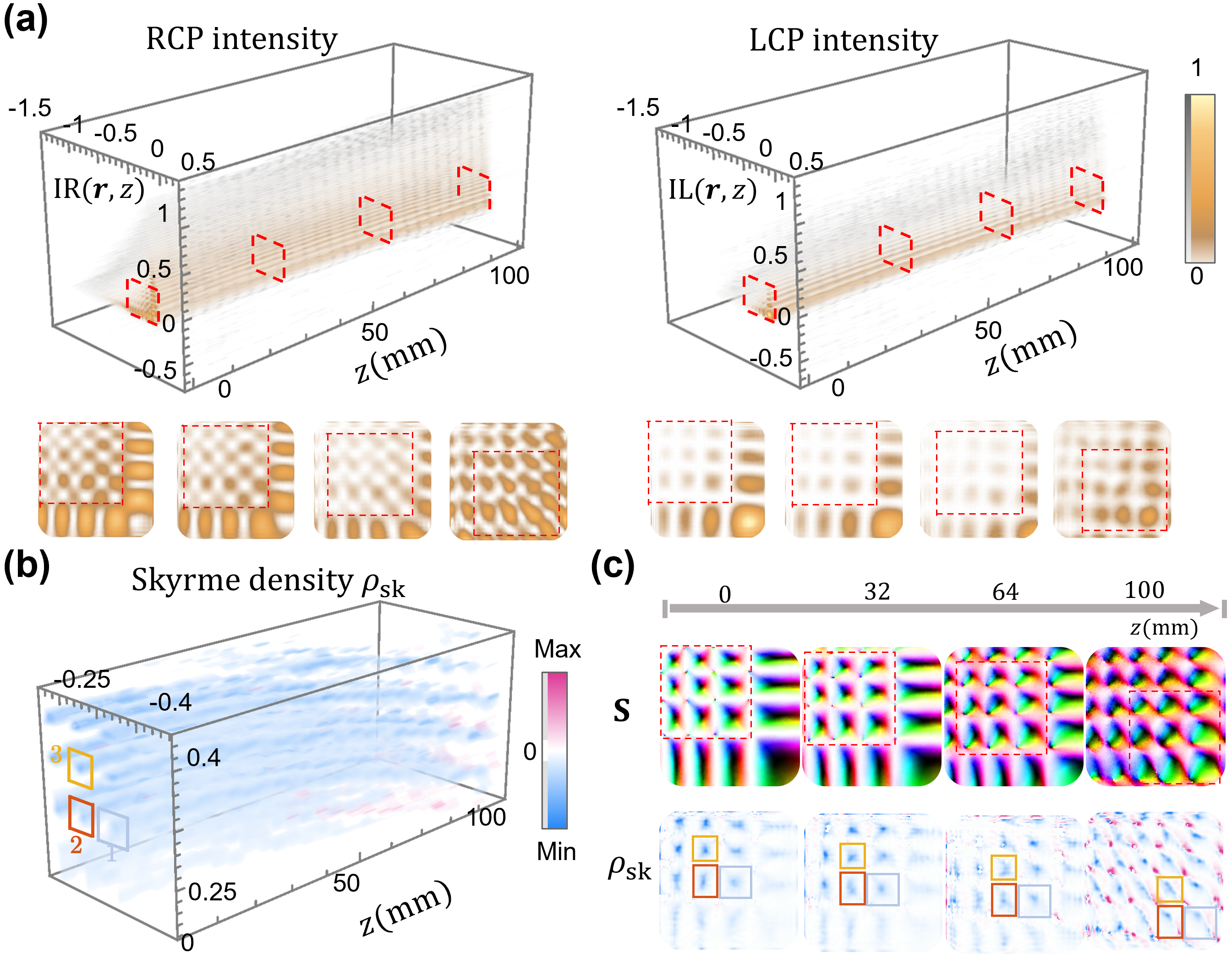}
\caption{\label{fig3}\textbf{Experimental results for periodically accelerating skyrmion lattices.} \textbf{a}, 3D intensity profiles of vortex Airy (RCP) and Airy (LCP) beams, reconstructed from experimental measurements, with selectively sampled transverse intensity lattices at different propagation distances (0 mm, 32 mm, 64 mm, and 100 mm). \textbf{b}, 3D skyrmion density distributions corresponding to the self-accelerating intensity lattices. \textbf{c}, Detailed distributions of the Stokes vector and skyrmion density at different z planes. As propagation distance increases, skyrmion textures degrade, and positive $\rho_{\mathrm{sk}}$ values (red) gradually emerge alongside the initially negative $\rho_{\mathrm{sk}}$ lattices (blue).}
\end{figure}

Fig.~\ref{fig3}(a) shows the observed 3D profiles of vortex Airy (RCP) and Airy (LCP) beams (reconstructed from 26 slices) with a Rayleigh length of $z_R=32.701$ mm, where both components propagate along diagonal curved trajectories (see \textbf{Supplementary Material S2} for more details). Transverse intensity profiles below the 3D profiles reveal periodic lattice structures that degrade with propagation, more prominently for vortex Airy beams. At $z=100$ mm, vortex lattices even become indistinguishable. The corresponding 3D skyrmion density distributions are presented in Fig.~\ref{fig3}(b). At the Fourier plane ($z=0$), perfectly periodic Stokes vector and skyrmion density distributions are observed (Fig.~\ref{fig3}(c)). However, as propagation distance increases, vortex lattice degradation leads to skyrmion cell deformation, a shrinking $\mathrm{S}_3=1$ area, and the emergence of abnormal positive $\rho_{\mathrm{sk}}$ values (Theoretical references for these results are provided in Fig.~S3). This raises an important question: how stable are the topological structures during propagation?

\begin{figure}
\includegraphics[width=\linewidth]{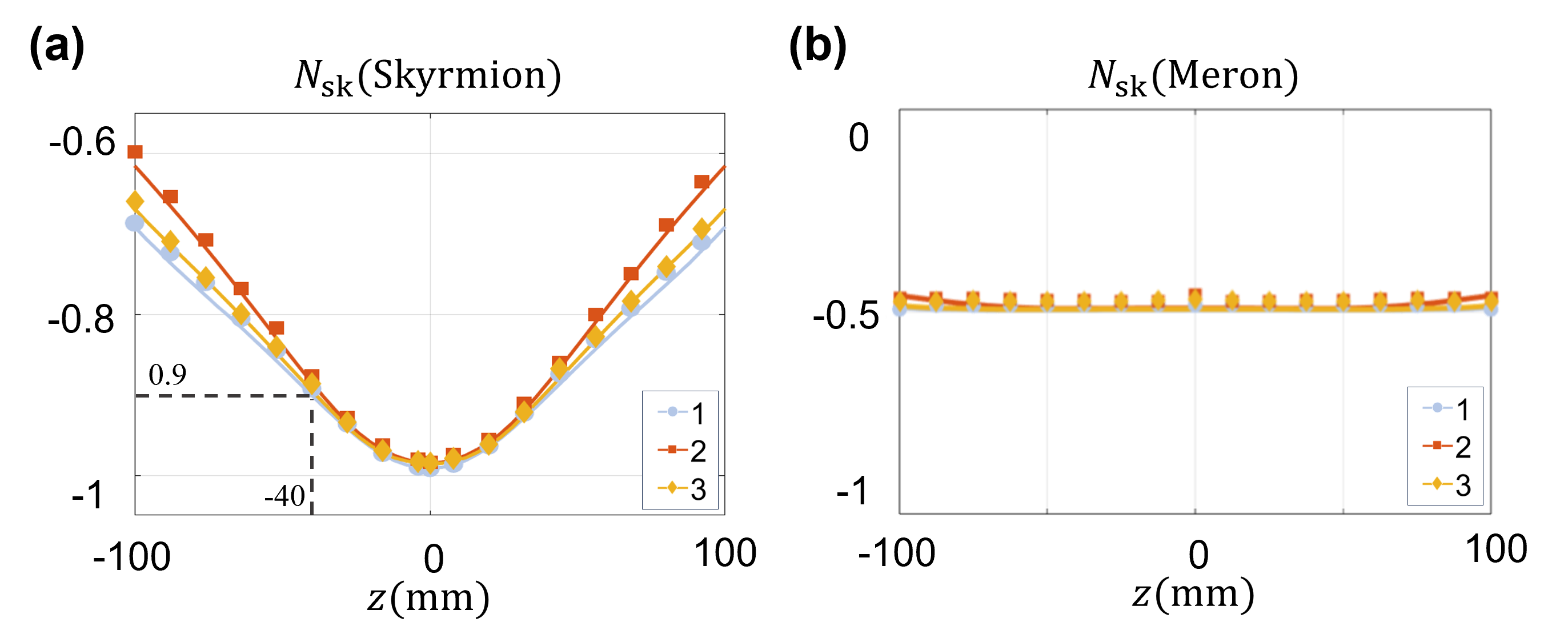}
\caption{\label{fig4}\textbf{Quantitative analysis of accelerating skyrmion and meron lattices.} \textbf{a}, Skyrme number for skyrmion lattices, showing a gradual decrease with propagation distance. \textbf{b}, skyrme number for meron lattices, remaining stable over propagation distance. Point data correspond to experimental observations, and solid lines represent theoretical results.}
\end{figure}

To quantify the stability, the skyrme numbers of three cells of skyrmion lattices (labeled 1, 2, and 3) were calculated, as shown in Fig.~\ref{fig4}(a). As the lattices degrade, the skyrme number $N_{\mathrm{sk}}$ gradually decreases with distance, with cell 2 showing the fastest decline due to its smaller integration area, which becomes insufficient for maintaining accuracy with increasing lattice complexity. Despite these challenges, skyrmion lattices maintain stable acceleration over a range of $\pm40$ mm ($\pm1.22\ z_R$), with $\left|N_{\mathrm{sk}}\right|>0.9$. Beyond this range, $N_{\mathrm{sk}}$ drops to approximately $\text{-}0.6$. 

Interestingly, meron lattices, which occupy the core regions of skyrmion structures, demonstrate a greater level of robustness. As shown in Fig. 4(b), meron lattices maintain stability across the entire measured range ($\pm3.06\ z_R$), with $N_{\mathrm{sk}}\approx-0.5$, as shown in Fig. ~\ref{fig4}(b). This stability is achieved at the expense of reduced  $\mathrm{S}_3=1$ regions, as the inner topological structures are better preserved.

Overall, Skyrmion lattices in this work not only inherit the stable acceleration and self-healing properties of Airy beams~\cite{chu2012analytical,broky2008self,zhang2015investigating} (see \textbf{Supplementary Material S3} for details of self-healing) but also introduce an additional degree of freedom due to their intricate topological structures. These features not only unlock new opportunities for exploring their dynamic behavior but also show promise for practical applications.

\section{Conclusion}
We have experimentally demonstrated a versatile family of optical skyrmion lattices capable of accelerating along curved trajectories in free space. These lattices are generated by imprinting cubic phase modulation onto traditional skyrmion beams. The transverse acceleration rate of these trajectories can be flexibly controlled by adjusting the scaling factor \textit{b}. Importantly, only LG beams that form skyrmion structures with odd topological charge differences can generate skyrmion lattices.

Leveraging the non-diffracting and self-accelerating nature of Airy beams, skyrmion lattices exhibit stable transverse acceleration and robust topological properties over a propagation range of $\pm1.22\ z_R$, even under propagation-induced lattice deformation. Greater robustness is observed for Meron lattices, which occupy the core regions of skyrmion structures and remain stable over an extended range of$\pm3.06\ z_R$. 

These unique features establish skyrmion lattices as a promising platform for exploring advanced applications such as particle sorting~\cite{baumgartl2008optically,schley2014loss,voloch2013generation}, advanced microscopy~\cite{schley2014loss,voloch2013generation,jia2014isotropic,vettenburg2014light,nylk2018light,wang2020airy}, plasma channel induction~\cite{schley2014loss,polynkin2009curved}, electron beam manipulation~\cite{schley2014loss,voloch2013generation}, and guiding electric discharges~\cite{clerici2015laser}, opening new frontiers in structured light research.

\section*{SUPPLEMENTARY MATERIAL}
See supplementary material for more details of Laguerre-Gaussian Modes, Airy and vortex Airy beams.

\begin{acknowledgments}
Singapore Ministry of Education (MOE) AcRF Tier 1 grant (RG157/23), MoE AcRF Tier 1 Thematic grant (RT11/23), and Imperial-Nanyang Technological University Collaboration Fund (INCF-2024-007), Nanyang Technological University Start Up Grant, and National Natural Science Foundation of China (12474324, 62075050, 11934013).

\end{acknowledgments}

\section*{Data Availability Statement}

The data that support the findings of this study are available from the corresponding author upon reasonable request.

\bibliographystyle{naturemag}

\bibliography{sample}

\end{document}